\documentclass[lettersize,journal]{IEEEtran}
\usepackage[table,xcdraw]{xcolor}
\usepackage{booktabs}
\usepackage{multirow}
\usepackage{graphicx}

\usepackage[skip=1ex]{caption}
\usepackage{array, booktabs, makecell, multirow}
\usepackage[utf8]{inputenc}
\usepackage{amsmath,amsfonts}
\usepackage{algorithmic}
\usepackage{algorithm}
\usepackage{array}
\usepackage[caption=false,font=footnotesize,labelfont=rm,textfont=rm]{subfig}
\usepackage{textcomp}
\usepackage{stfloats}
\usepackage{url}
\usepackage{verbatim}
\usepackage{graphicx}
\usepackage{caption}
\usepackage{cite}
\usepackage{doi}
\usepackage{multirow}
\usepackage[table,xcdraw]{xcolor}
\usepackage{amssymb}
\usepackage{pifont}
\usepackage{tcolorbox}
\tcbuselibrary{most}
\usepackage{booktabs}
\DeclareUnicodeCharacter{FF08}{(}
\DeclareUnicodeCharacter{FF09}{(}
\begin{document}

\title{SoK: Detection and Repair of Accessibility Issues}
\author{
Liming Nie,
Hao Liu,
Jing Sun*,
Kabir Sulaiman SAID,
Shanshan Hong,
Lei Xue,
Zhiyuan Wei, 
Yangyang Zhao,
and Meng Li* 
\thanks{L. Nie, H. Liu, S. Hong and M. Li are with Shenzhen Technology University, Shenzhen, China}
\thanks{J. Sum is with Jiliang University, Hangzhou,China}
\thanks{K. Sulaiman SAID is with Aliko Dangote University of Science and Technology, Wudil, Kano, Nigeria}
\thanks{Z. Wei is with University of Chicago, Chicago, USA.}
\thanks{Y. Zhao is with Zhejiang Sci-Tech University, Hangzhou, China.}
\thanks{L. Xue is with Sun Yat-Sen University, Shenzhen, China}

}




\maketitle

\begin{abstract}
There is an increasing global emphasis on information accessibility, with numerous researchers actively developing automated tools to detect and repair accessibility issues, thereby ensuring that individuals with diverse abilities can independently access software products and services. However, current research still encounters significant challenges in two key areas: the absence of a comprehensive taxonomy of accessibility issue types, and the lack of comprehensive analysis of the capabilities of detection and repair tools, as well as the status of corresponding datasets.
To address these challenges, this paper introduces the Accessibility Issue Analysis (AIA) framework. Utilizing this framework, we develop a comprehensive taxonomy that categorizes 55 types of accessibility issues across four pivotal dimensions: Perceivability, Operability, Understandability, and Robustness. This taxonomy has been rigorously recognized through a questionnaire survey (n=130).
Building on this taxonomy, we conduct an in-depth analysis of existing detection and repair tools, as well as the status of corresponding datasets. In terms of tools, our findings indicate that 14 detection tools can identify 31 issue types, achieving a 56.3\% rate (31/55). Meanwhile, 9 repair tools address just 13 issue types, with a 23.6\%  rate. In terms of datasets, those for detection tools cover 21 issue types, at a 38.1\% coverage rate, whereas those for repair tools cover only 7 types, at a 12.7\% coverage rate.
\end{abstract}


\section{Introduction}
\begin{figure}[htbp]
\centering
\includegraphics[width=1\linewidth]{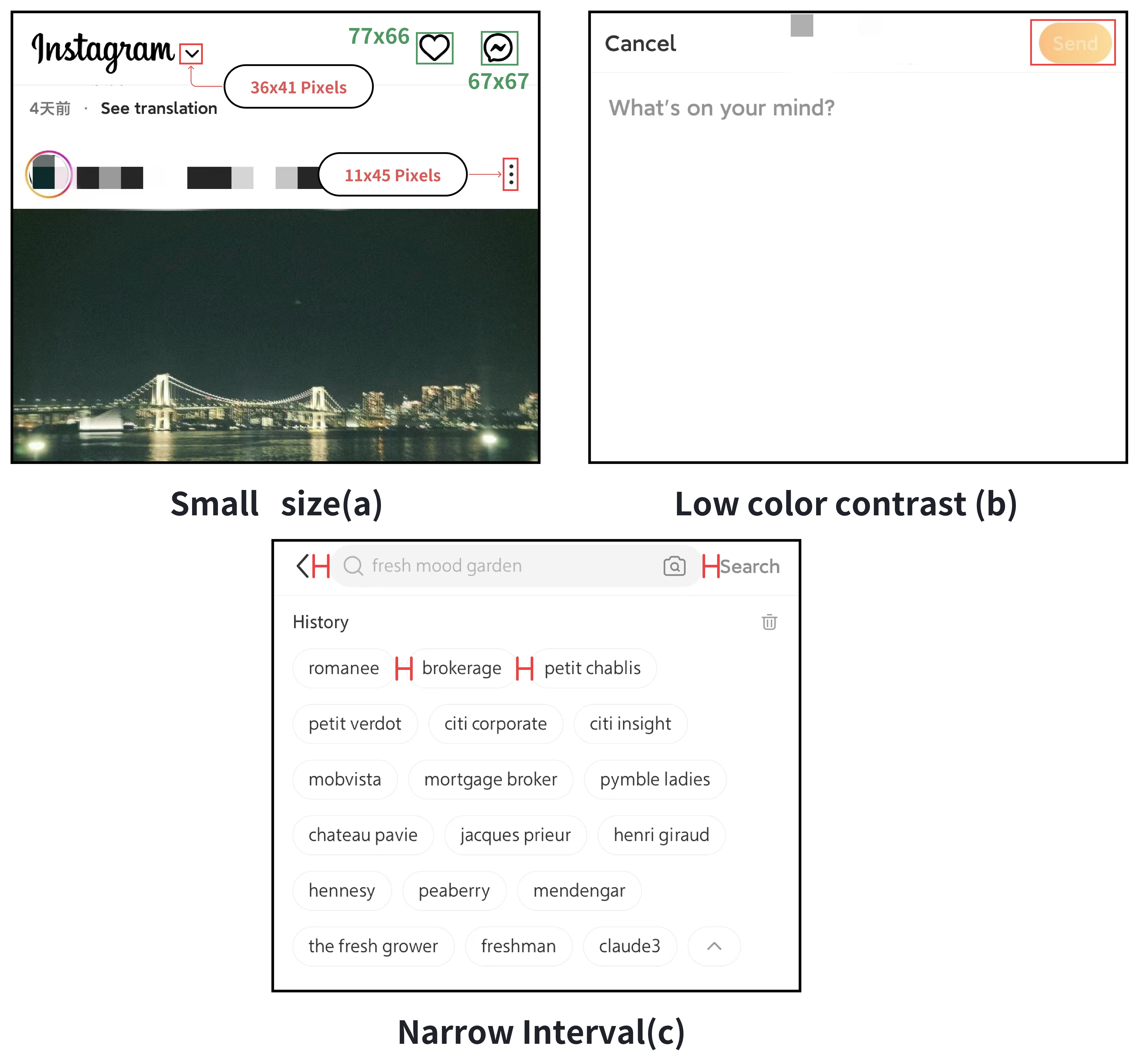}
\caption{The examples of accessibility issues} \label{accessibility issues}
\end{figure}

\IEEEPARstart{W}{ith} the rapid advancement of information technology, accessibility considerations are garnering growing attention within the realm of software engineering\cite{45}\cite{46}. However, the design and implementation of current software and websites frequently neglect the needs of people with disabilities, posing substantial barriers to their access to information and social participation \cite{44}.For instance, small touch targets in user interfaces—such as buttons and icons—as well as icons positioned too closely together(in Fig \ref{accessibility issues}.a,c), impede users with motor impairments from engaging in effective interaction\cite{3},\cite{15},\cite{13},\cite{4},\cite{50},\cite{51}. Similarly, insufficient contrast between text and background colors (below 3.0) (in Fig \ref{accessibility issues}.b)complicates reading and information access for those with visual impairments\cite{5},\cite{11},\cite{43}. 

To address pervasive accessibility issues in mobile applications and websites, researchers have leveraged a variety of technologies, including large language models, convolutional neural networks, natural language processing, among others, to create automated tools for detecting and repairing accessibility issues\cite{8}, \cite{9}, \cite{10}, \cite{11}, \cite{28}, \cite{12}, \cite{13}, \cite{3}, \cite{38}, \cite{14}, \cite{15}, \cite{16}, \cite{17}, \cite{18}, \cite{19}, \cite{20}, \cite{21}, \cite{22}. Specifically, In the realm of detection, Krishnavajjala et al. \cite{3} developed MotorEase, a tool that integrates the Faster-RCNN model, computer vision, and text processing to automatically identify accessibility issues impacting users with motor impairments in mobile app interfaces. Taeb et al. \cite{39}  created AXNav, a system leveraging large language models and pixel-based UI understanding models to transform natural language accessibility test instructions into reproducible, navigable videos for issue identification and testing. He et al. \cite{28} created AdMole, an automated tool based on Groundhog for accessibility evaluation, which leverages UI screenshot analysis and ad element recognition to detect accessibility issues specifically in ads within Android applications. In the area of repair, Zhang et al. \cite{15} proposed AccessFixer, a method utilizing a graph convolutional neural network to automatically modify GUI component properties for enhanced accessibility. Zhang et al. \cite{14} developed Iris, a tool integrating automation and context-awareness technologies to tackle color-related accessibility issues in Android apps. Alotaibi et al. \cite{16} proposed the SALEM , which combines the SRG model with genetic algorithms to fix small touch target issues in mobile applications.

Despite notable advancements, current research confronts several challenges. Firstly, existing studies have not thoroughly collected and classified the accessibility issues users face when using mobile applications and websites.A comprehensive taxonomy for accessibility issue types would enhance understanding of these issues and aid developers in crafting more accessible software applications \cite{50}.Secondly, the lack of integrated analyses of detection and repair tools capabilities and the current status of related datasets.The effectiveness of detection and repair tools directly influences the efficiency of issue identification and resolution, yet no study has analyzed their capabilities within a comprehensive taxonomy. Furthermore, high-quality datasets are crucial for tool performance \cite{49}. Despite the proposal of some datasets, research on their current status remains inadequate.

In this paper, we propose for the first time the Accessibility Issue Analysis Framework (AIA), a novel multistage approach designed to address these challenges. In the first stage, we systematically gather literature on accessibility issues through a rigorous literature review and snowballing method. In the second stage, based on the accessibility issue types collated and de-duplicated from these studies, we develop a comprehensive taxonomy, classifying each type according to WCAG 2.1. Additionally, we provide detailed annotations for each issue type, including the impacted user groups and potential application scenarios. In the third and fourth stages, leveraging the constructed taxonomy, we evaluate the capabilities of existing detection and repair tools and analyze the current status of datasets used for accessibility issue detection and repair. 

Based on this framework, we answer three key research questions (RQs):
\begin{itemize}
\item \textbf{RQ1: What known accessibility issues do users encounter when using mobile applications or websites?} 

We have developed the most comprehensive taxonomy to date, encompassing 55 recognized accessibility issue types, recognized through an questionnaire survey. Each type is annotated with the impacted user groups and potential application scenarios, enhancing the taxonomy’s applicability.
\item \textbf{RQ2: What are the capabilities of current tools for detecting and repairing accessibility issues?}

Of the 55 issue types, the 14 detection tools can identify 31 types, achieving a 56.3\% rate. Meanwhile, the 9 repair tools can address only 13 types, with a 23.6\% rate. This indicates significant gaps in the capabilities of current tools.
\item \textbf{RQ3: What is the status of datasets used for detection and repair tools in relation to the taxonomy?}

Currently, the 10 datasets for detection tools encompass 21 out of the 55 types of accessibility issues, achieving a coverage rate of 38.1\%. Additionally, the 8 datasets for repair tools address only 7 issue types, resulting in a coverage rate of 12.7\%. The lack of comprehensive datasets limits the effectiveness of current detection and repair tools.
\end{itemize}

By answering these questions, this paper enhances our comprehension of accessibility issue type classification, as well as the capabilities of detection and repair tools, thereby providing valuable insights for the field’s future research and practice.This paper makes the following contributions:
\begin{itemize}
\item Constructed a comprehensive taxonomy  for accessibility issue types, providing a standardized reference framework for future research.
\item Analyzed the capabilities of current detection and repair tools and revealed their limitations.
\item Evaluated the current status of datasets for detection and repair tools and provided guidance for future dataset development.
\end{itemize}

\section{Background and Related Work}
In this section, we briefly describe the prevalent accessibility issues in mobile applications and websites, the Web Content Accessibility Guidelines 2.1, the current research on accessibility issue detection and repair, and existing reviews of accessibility issues.

\subsection{Accessibility Issues in Mobile Applications and Websites}
Alshayban et al. \cite{5} analyzed over 1,000 Android apps and found that almost all apps are riddled with accessibility issues, hindering their useby disabled people..By 2019, approximately 70\% of websites online contained accessibility barriers, which restricted individuals with disabilities from accessing key features and limited their equal engagement in the digital realm \cite{26}. In February 2024, WebAIM evaluated the accessibility of more than one million website homepages, discovering that 95.9\% contained accessibility issues, averaging 56.8 errors per page and totaling 56,791,260 errors.With the increasing complexity of web elements, it is estimated that users with disabilities face accessibility issues with approximately 21 elements per page \cite{32}. Being able to easily access and use information is important for everyone, especially for the approximately 15\% of people with disabilities worldwide \cite{25}. Therefore, improving the accessibility of digital environments is particularly necessary.

\subsection{Web Content Accessibility Guidelines 2.1}
The W3C's Web Accessibility Initiative (WAI) developed the WCAG to make web content more accessible to individuals with disabilities.WCAG 2.1\cite{24}, the most recent international standard published by the W3C, is founded on four key principles: perceivable, operable, understandable, and robust.These principles guarantee that content is accessible, operable, and understandable for all users and can be consistently presented across different user agents and assistive technologies.WCAG 2.1 provides extensive guidance for creating accessible designs.

\subsection{Detection and Repair of Accessibility Issues}
In recent years, researchers have proposed diverse tools and algorithms to enhance the identification and repair of accessibility issues\cite{8}, \cite{9}, \cite{10}, \cite{11}, \cite{28}, \cite{12}, \cite{13}, \cite{3}, \cite{38}, \cite{14}, \cite{15}, \cite{16}, \cite{17}, \cite{18}, \cite{19}, \cite{20}, \cite{21}, \cite{22}. 

In terms of detecting accessibility issue, researchers have proposed multiple tools that often leverage techniques such as machine learning, deep learning, dynamic program analysis, and heuristic rules. For instance, Alshayban et al. \cite{8} developed AccessiText, an automated testing technique that combines dynamic analysis with heuristic rules to specifically detect text accessibility issues in Android applications. Chiou et al. \cite{9} analyzed different approaches to keyboard navigation of web UIs for keyboard users and constructed a keyboard dialog model to detect navigation failures in websites. Salehnamadi et al. \cite{10} designed the automated accessibility testing tool Groundhog, which uses three different agents to simulate user interactions and detect accessibility issues in mobile applications. Tazi et al. \cite{11} developed a tool based on Accessibility Insights to detect common accessibility issues through user interface analysis. He et al. \cite{28} created AdMole, an accessibility evaluation tool built on Groundhog, that employs UI screenshot analysis and ad element recognition to specifically detect accessibility issues in ads within Android applications. Zhang et al. \cite{12} proposed a deep learning-based method that infers accessibility metadata by analyzing UI pixels. Salehnamadi et al. \cite{13} combined dynamic program analysis with Accessibility Scanner and the Appium testing framework to evaluate the accessibility of GUI elements through automated testing scripts. Krishnavajjala et al. \cite{3} developed MotorEase, which integrates Faster-RCNN, computer vision, and text processing techniques to automatically detect accessibility issues affecting motor-impaired users in mobile app UIs. Bajammal and Mesbah \cite{38} introduced AXERAY, which uses hierarchical visual analysis to infer the semantic grouping and roles of webpage elements for automated accessibility testing.

In terms of repairing accessibility issues, researchers have used graph models, convolutional neural networks, natural language processing, and genetic algorithms for issue repair. Specifically, Zhang et al. \cite{14} developed the Iris tool, which combines automated and context-aware technology to address color-related accessibility issues in Android applications. Zhang et al. \cite{15} proposed the AccessFixer method, which uses a relational graph convolutional neural network model to automatically adjust GUI component attributes. Alotaibi et al. \cite{16} proposed the SALEM tool, which combines the SRG model with genetic algorithms to repair issues of touch target size in mobile apps. Mehralian et al. \cite{17} proposed the coala method for generating accurate icon labels. Xu et al. \cite{18} proposed the AGAA method, which converts GUIs into graph structures and uses genetic algorithms to generate accessibility issues on Android apps. Chen et al. \cite{19} developed the LabelDroid method, which automatically predicts natural language labels for GUI components. Zhang et al. \cite{20} proposed the SAM tool, which supplements missing alternative text in SVG buttons. Alotaibi et al. \cite{21} developed the ScaleFix tool, which automatically repairs user interface scaling accessibility issues. Zhang et al. \cite{22} proposed the Screen Recognition tool, which combines heuristic rules and OCR technology to create accessibility metadata from pixels.

In summary, although numerous studies have focused on detecting and repairing accessibility issues, there is still a significant lack of research analyzing the capabilities of detection and repair tools.Building on this, this paper collects and reviews these detection and repair tools and comprehensively evaluates their capabilities based on the taxonomy we developed.

\subsection{Existing Reviews of Accessibility Issues}
Beyond the aforementioned research on accessibility issue detection and repair, several studies have focused on comprehensive reviews of accessibility issues. These reviews can be broadly categorized into two main areas: those that concentrate on specific disability groups, such as visually impaired individuals, in their use of digital technologies \cite{72},\cite{73},\cite{74},\cite{75},\cite{76} and the other that explores accessibility issues faced by individuals with disabilities in specific contexts \cite{77},\cite{78},\cite{79},\cite{80},\cite{81}.

Specifically, Kerdar et al. \cite{72} conducted a systematic review of 49 studies from 2004 to 2019, delving into digital accessibility issues through the firsthand experiences and challenges of individuals with visual impairments and blindness. Agrimi et al. \cite{73} focused their review on the accessibility of games for visually impaired individuals, finding that while some games are specifically designed for this group, the majority do not adequately consider their unique needs, leading to numerous barriers within the gaming environment. 
Khalajzadeh et al. \cite{74} reviewed 38 studies from 2004 to 2021, aiming to explore accessibility issues in low-code approaches and noting that users with visual impairments face various challenges when using low-code platforms. 
Moreno et al. \cite{75} conducted a systematic literature review to explain the reasons why accessibility barriers for the elderly remain unresolved and to identify areas needing further effort. 
Borina et al. \cite{76} reviewed 44 studies from 2015 to 2021 to assess web accessibility for individuals with cognitive impairments, revealing that most websites have poor accessibility.
Dai et al. \cite{77} examined 30 academic publications to analyze the challenges faced by elderly individuals and those with neurodiverse needs when using online banking; Mountapmbeme et al. \cite{78} analyzed 70 papers to identify programming accessibility barriers encountered by visually impaired individuals learning to code; 
Nevsky et al. \cite{79} systematically reviewed 181 papers from 1996 to 2022, exploring the primary accessibility challenges faced by individuals with disabilities when accessing digital audiovisual media and the focus of audiovisual media accessibility research; 
Deriba et al. \cite{80} analyzed 21 papers to investigate accessibility barriers in virtual laboratories and explore potential solutions; 
M. Akram and R. Bt Sulaiman \cite{81} conducted a systematic literature review of 15 studies within and outside Saudi Arabia, examining web accessibility issues in government and university websites.

Despite providing valuable insights, these review studies have largely focused on specific disability groups or specific contexts. In contrast, this paper systematically reviews the relevant literature to explore a broader spectrum of accessibility issues encountered by various disability groups, including those with visual, motor, hearing, cognitive impairments, and aphasia, across diverse contexts. Additionally, based on the WCAG 2.1 guidelines, we have categorized the identified issues and developed a comprehensive taxonomy. Utilizing this taxonomy, we conducted an in-depth analysis of the capabilities and limitations of current accessibility issue detection and repair tools, as well as assessed the status of corresponding datasets. This paper aims to fill the gaps in existing research, providing a more comprehensive reference and guidance for the further development of accessibility technologies.

\begin{figure}[htbp]
\centering
\includegraphics[width=1\linewidth]{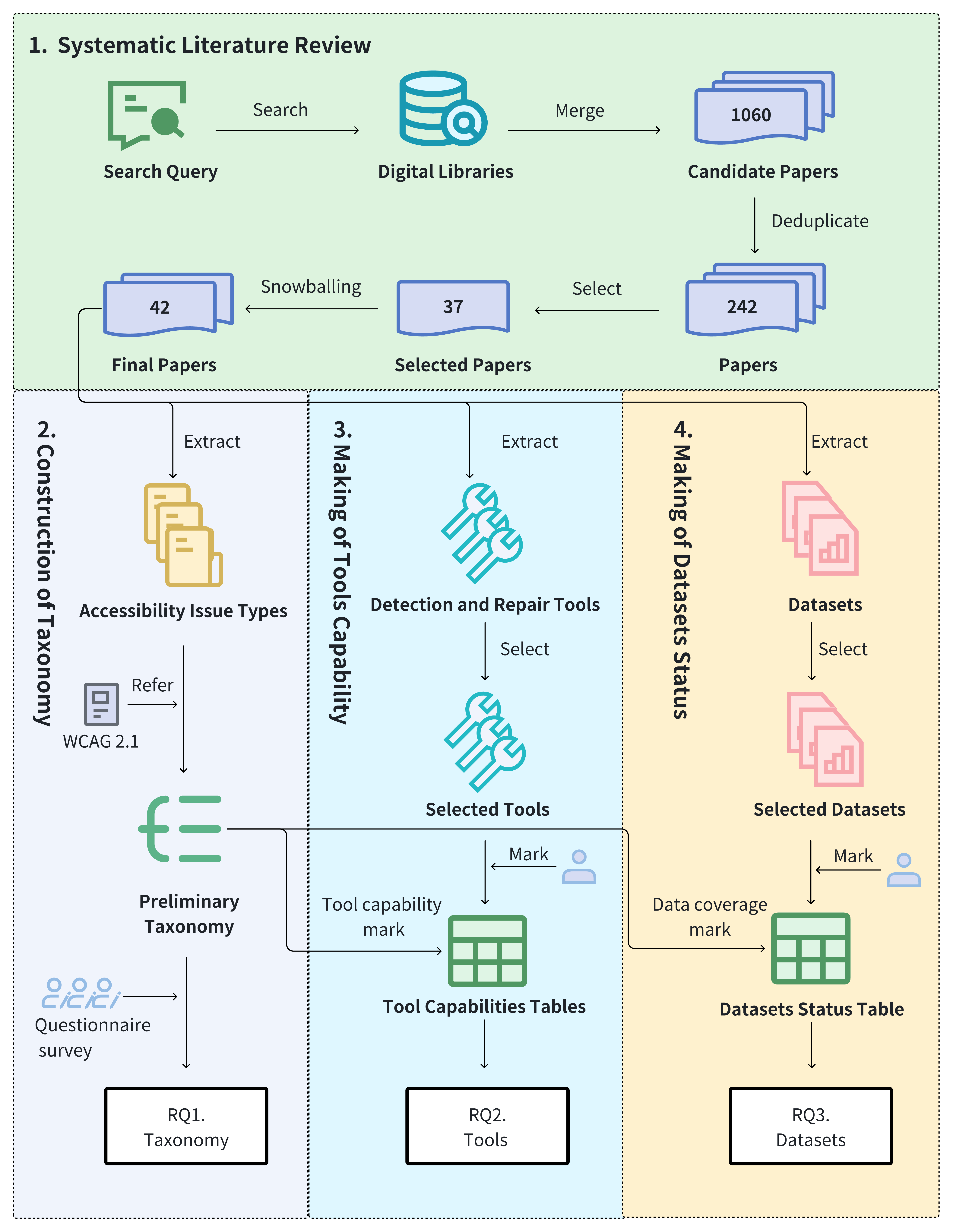}
\caption{The framework of AIA} \label{Framework}
\end{figure}

\section{Methodology}
This section begins by outlining the three research questions and their underlying motivations. Following this, we introduce a detailed framework intended to direct our empirical investigation in addressing these questions.

\subsection{Research Questions And Motivation}
\begin{itemize}
\item \textbf{RQ1: What known accessibility issues do users encounter when using mobile applications or websites?} 

To effectively understand, detect, and repair accessibility issues, a comprehensive and widely accepted taxonomy of accessibility issue types is necessary. However, research on developing such a taxonomy is relatively limited. At present, such a taxonomy does not exist. Consequently, constructing a comprehensive taxonomy of accessibility issues is of paramount importance.
\item \textbf{RQ2: What are the capabilities of current tools for detecting and repairing accessibility issues?}

Although several automated tools for detecting and repairing accessibility issues have been developed, their capabilities have not yet been thoroughly examined. An in-depth analysis of these tools’ capabilities is crucial for practitioners to select the most suitable ones and for future optimization and expansion of their application scope.
\item \textbf{RQ3: What is the status of datasets used for detection and repair tools in relation to the taxonomy?}

One major obstacle in research related to accessibility issue detection and repair is the lack of comprehensive datasets. Although some datasets exist for use in detection and repair tools, their status has not been adequately investigated, particularly regarding their actual coverage of a comprehensive accessibility issue types taxonomy. Thus, an in-depth analysis of existing datasets is both necessary and urgent.
\end{itemize}

\subsection{Framework}
\label{section:Framework} 
As shown in Figure \ref{Framework},to address the three research questions mentioned above, we propose an Accessibility Issue Analysis Framework (AIA), based on the taxonomy construction approach used by Ladisa et al. \cite{40} and the standardization methods proposed by Usman et al. \cite{6} and Ralph et al. \cite{7}. It aims to construct a comprehensive taxonomy of accessibility issue types and, based on this taxonomy, provide in-depth analyses of the capabilities of existing tools and the status of related datasets. AIA includes four modules: systematic literature review, taxonomy construction, tool capability analysis, and dataset status analysis. First, the systematic literature review module comprehensively collects and screens literature on accessibility issues to provide a theoretical foundation for taxonomy construction. Next, the taxonomy construction module extracts accessibility issue types from the screened literature and builds a comprehensive taxonomy, providing a reference for the analysis of tools and datasets. Third, the tool capability analysis module selects tools for detecting and repairing accessibility issues and evaluates their capabilities based on the taxonomy. Finally, the dataset status analysis module assesses the current status of datasets used for detection and repair tools in relation to the taxonomy.


\begin{figure}[htbp]
\centering
\includegraphics[width=1\linewidth]{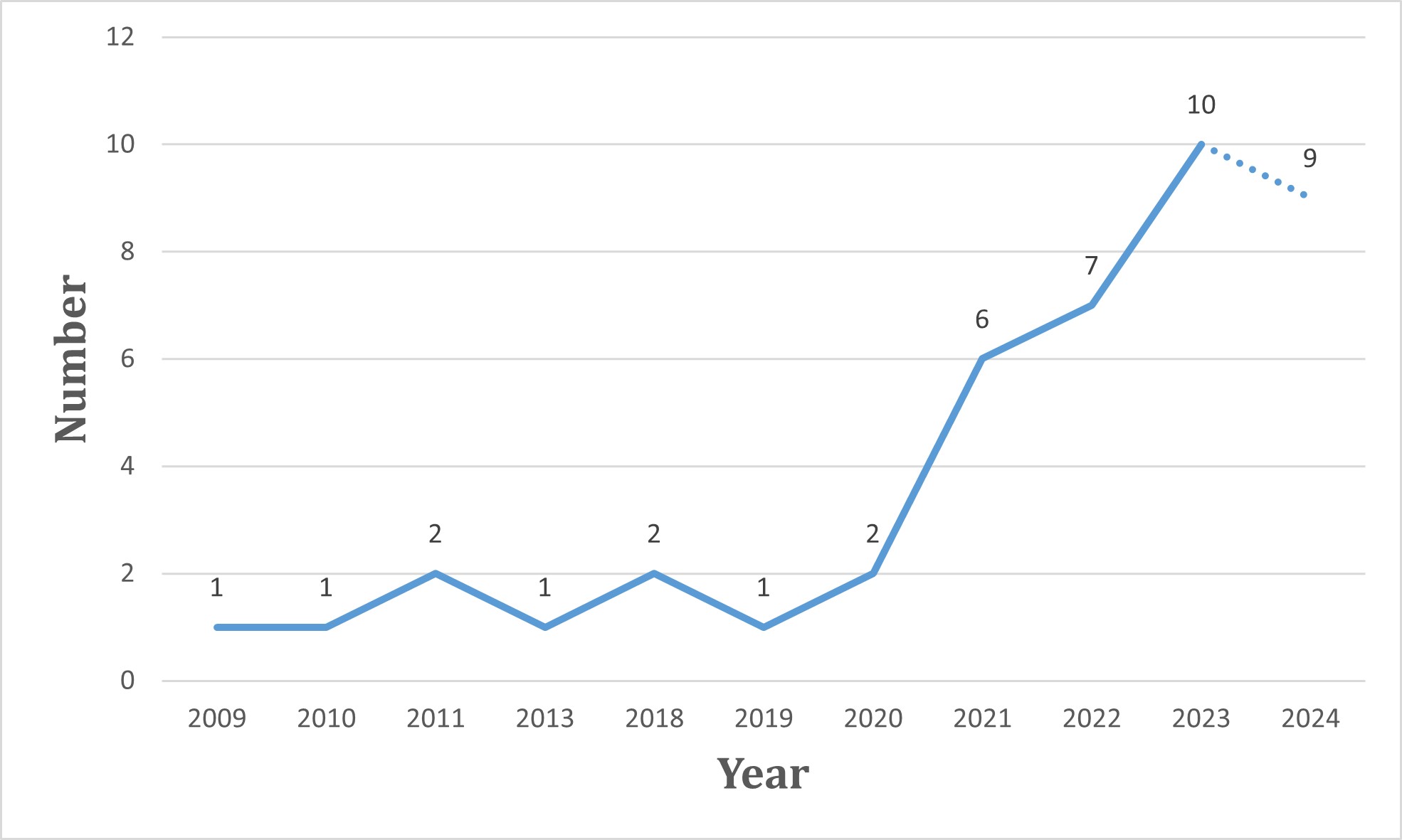}
\caption{Annual Publication Trends of Acccessibility Issue-Related Research Papers} \label{annual paper}
\end{figure}

\subsection{Systematic Literature Review}
\label{section:Systematic Literature Review}
This subsection introduces the collection of literature related to accessibility issue research, as shown in the first module of Figure \ref{Framework}. To achieve this goal, we followed a structured approach for the systematic literature review \cite{41},\cite{42}, comprising two stages: candidate literature search and relevant literature screening.

\subsubsection{Candidate Literature Search}
To conduct an effective candidate literature search, we performed the following three steps: search query design, actual search execution, and deduplication of retrieved papers.

Our systematic literature review on accessibility issue research began with designing an effective search query. To ensure a robust query, we conducted an initial exploratory search using the keyword ``Accessibility Issues" in the ACM Digital Library. This initial search helped us identify several related studies \cite{3},\cite{4},\cite{5},\cite{20},\cite{28},\cite{29},\cite{30},\cite{31}, which provided a preliminary understanding of the topic and assisted in identifying a range of keywords related to accessibility issues. Subsequently, we analyzed these keywords and constructed a search query aimed at retrieving literature pertinent to accessibility issues. The search query is as follows: (Accessibility OR Accessibility issue OR Assistive Technology OR Mobile accessibility OR Accessibility issue repair OR Color-related accessibility issue OR accessibility guidelines OR blind and visually impaired OR web accessibility OR WCAG).

Using this carefully crafted search query, we conducted a systematic literature search across five authoritative digital libraries: Google Scholar, IEEE Xplore, ACM Digital Library, Science Direct and DBLP. These libraries were selected for their extensive and high-quality academic resources in computer science and software engineering, ensuring the comprehensiveness and authority of the literature search. During the search process, to maintain result relevance, we set the following criteria: search scope was limited to the title, abstract, and keywords of papers, restricted to English, and included only academic journals and conference papers. The entire search process was conducted manually over two weeks, ultimately yielding 1060 candidate papers. It is important to note that duplicates may exist among these papers as they were retrieved from different digital libraries.

Lastly, to eliminate duplicate papers, we performed a rigorous manual deduplication process. First, we exported the information of all retrieved papers into a table, where each entry represented a document and included title, author list, publication year, and abstract information. Then, the second author of our team sorted the table by title and conducted a line-by-line manual check, marking and removing duplicate entries. After this careful deduplication process, we ultimately compiled a set of 242 unique papers.

\begin{table}
\caption{Conference Venues and Journal Venues List}
\label{Conference Venues and Journal Venues List}
\resizebox{1.0\linewidth}{!}{
\begin{tabular}{ll} 
\toprule
\textbf{Conference Venues} & \begin{tabular}[c]{@{}l@{}}ICSE*,~ FSE*, ASE*, ICSME*, MSR*, ICPC*, ISSTA*, \\ICST*, SANER*, UIST*, CHI*, SPLASH*, OOPSLA*, \\PLDI*, CSCW*, ASSETS*, USS, ICMI, COMPSAC\end{tabular}  \\ 
\midrule
\textbf{Journal Venues}    & \begin{tabular}[c]{@{}l@{}}TSE*, TOSEM*,EMSE*, JSS*, ASE*, Soft Computing,\\MTA, BIT, SOCA, WWW, AEJ, SCP, IJHCS, ESWA\end{tabular}      \\
\bottomrule
\end{tabular}}
\vspace{-1.5em}
\end{table}

\subsubsection{Relevant Literature Screening}
To identify the most relevant papers, we implemented a four-step screening process: defining the literature scope, preliminary screening based on metadata, a detailed full-text review, and applying the snowball technique to the screened literature.

Firstly, to define the literature scope, we identified 19 conferences and 14 journals, as listed in the Table \ref{Conference Venues and Journal Venues List}. Journals and conferences marked with an asterisk (*) in the table were taken from the study by Vajjala et al. \cite{3}. In addition, we included three more conferences (USS, ICMI, COMPSAC) and nine more journals (such as MTA, Soft Computing, and BIT) to cover a broader range of literature related to accessibility research. We limited the 242 deduplicated papers to the scope of these conferences and journals, ultimately reducing the candidate literature to 73 papers.

Secondly, to ensure that the selected papers was directly related to accessibility issue research, we conducted a preliminary screening based on metadata. In this step, we carefully examined the publication type, title, abstract, and keywords of each document, with particular attention to its relevance to accessibility issue research. Any papers clearly unrelated to accessibility issue research were excluded. This preliminary screening reduced the candidate literature to 67 papers.

Next, to ensure that the selected papers encompasses adequate information for our subsequent review, we obtained the full text of the screened papers for a final review. We focused on the introduction, methodology, and results sections to ensure that these papers were not only related to accessibility issue research but also provided detailed descriptions of the accessibility issues, such as their descriptions, occurrence scenarios, and affected populations. Additionally, all sections of the study had to be written in English to be considered. Papers not meeting these standards were excluded. After this round of rigorous screening step, the number of candidate literature was narrowed down to 37.

Finally, to gather as comprehensively as possible accessibility issue research,we applied a bidirectional snowball strategy to the 37 papers selected in the previous step. This approach involved tracing both forward and backward references, focusing exclusively on papers that directly cited or were cited by the selected papers, and adhering to a one-layer depth to maintain focus and control over the results. During this systematic supplementary search, we meticulously screened any additional literature found according to the same selection criteria. As a result, we identified and incorporated 5 new papers, bringing the total to 42 relevant papers.

The Figure \ref{annual paper} shows the publication trend of the 42 selected papers by year. Based on our analyses, it is evident that research on accessibility issues has been gaining substantial momentum in recent years. From 2011 onwards, the number of studies in this field remained relatively low, with only one to two papers published annually. However, a significant growth trend emerged starting in 2021. Specifically, the volume of research increased from six papers in 2021 to seven in 2022, and reached a peak of 10 papers in 2023. Notably, as of July 2024, nine studies have been published, further highlighting the growing interest and sustained investment in this domain.

\subsection{Construction of the Taxonomy}
\label{section:Construction of the Taxonomy}
This section outlines the methodology for constructing the taxonomy of accessibility issue types, as well as the questionnaire survey on the taxonomy, as shown in the second module of Figure \ref{Framework}.
\subsubsection{Construction}
We adopted a multistep approach to constructing the taxonomy, which included three main steps: collection of accessibility issue types, merging of similar types guided by the WCAG 2.1, and categorization of issue types based on the WCAG 2.1.

First, regarding the collection of accessibility issue types, we followed these steps: we designed a structured table covering key information such as the issue type description, impact, affected user group, and occurrence platform. Then, we conducted a thorough full-text review of the 42 selected relevant papers and manually recorded detailed annotations in the table. Ultimately, we identified and extracted 124 clearly described accessibility issue types from these papers, which provided a foundation for our subsequent classification and in-depth analysis. It is worth noting that we observed redundancy in accessibility issue types across different papers during the extraction process.

Second, to eliminate redundancy, we consolidated similar accessibility issue types. Specifically, we conducted a thorough review of all accessibility issue types, identifying redundant ones that often shared similarities in their descriptions, manifestations, or affected user groups (e.g., visual impairments, hearing impairments). These potentially redundant types were then categorized and flagged for further examination. We meticulously assessed these types, paying particular attention to whether they violated the same WCAG 2.1 success criteria. This entire process was carried out manually, with two authors independently evaluating each issue type based on textual descriptions, violated success criteria, and affected user groups. Any discrepancies were resolved through impartial arbitration by a third author. Ultimately, we merged issue types that demonstrated high consistency in their definitional descriptions, manifestations, affected user groups, and violations of WCAG 2.1 success criteria, retaining one issue description while consolidating other relevant information into a single entry. As a result, we successfully reduced the initial 124 accessibility issue types to 55.

Lastly, we categorized the deduplicated issue types based on WCAG 2.1. Specifically, we first thoroughly reviewed the four major principles of the WCAG 2.1 and their corresponding success criteria. Then, we examined the deduplicated accessibility issue types and assigned them to appropriate categories based on the WCAG 2.1 criterion each issue violated. For example, the issue "Text and image color contrast is lower than 4.5:1" violates WCAG 2.1’s Perceivable principle under criterion 1.4.3, Color Contrast (Minimum), so we categorized it under Perceivable. Issues not explicitly covered by WCAG 2.1 were assigned to appropriate categories based on their actual impact on users; for example, "Icons are too close together" affects user operation and was thus assigned to the Operable category. Through this approach, we developed a comprehensive taxonomy of accessibility issue types.

With these essential steps successfully completed, we have constructed a taxonomy of accessibility issue types that is as comprehensive as possible. This taxonomy encompasses known accessibility issue types in current academic research, aiming to provide a common reference framework for researchers, developers, policymakers, and other stakeholders in the accessibility field to more effectively identify, discuss, and resolve accessibility issues.

\subsubsection{Questionnaire Survey}
The goal of this survey is to gather feedback from a wide array of societal individuals regarding our developed taxonomy of accessibility issue types. The survey focuses on four key areas: a). Evaluating the taxonomy’s rationality, completeness, and understandability. b). Assessing the appropriateness of annotations for "people affected" and "WCAG 2.1." c). Evaluating the effectiveness of existing tools for detecting and repairing accessibility issues. d). Assessing the usability of existing accessibility issue datasets.

The survey consists of three parts:The first part focused on questionnaire design. The questionnaire includes 14 questions, each with a specific evaluation objective. Of these, 8 questions use a Likert scale ranging from 1 (lowest) to 5 (highest) to quantify participant feedback. There are also two binary questions (yes/no), such as whether participants reviewed the provided taxonomy document and background materials, and whether they are aware of accessibility issues in their applications. Additionally, four demographic questions were included, such as age, years of work experience,
occupational and industry background, to enable analysis of perspective differences across various backgrounds.The second part covered questionnaire distribution and scoring methods. To ensure broad representation, the survey was distributed to a diverse group. Participants were required to read the taxonomy document and background materials before scoring each question to ensure accuracy in their feedback.The third part focused on privacy protection. We prioritized protecting participants' privacy and anonymity. No personally identifiable information was collected during the survey, and we strictly adhered to data minimization principles. All data was securely stored, and only aggregate data was presented in the report to protect individual privacy.

\begin{table*}
\vspace{-1.5em}
\caption{Taxonomy of Accessibility issue types}
\label{Taxonomy}
\centering
\renewcommand{\arraystretch}{1.2}
\resizebox{1.0\linewidth}{!}{
\begin{tabular}{lllllcll}
\toprule
\textbf{Categories}  & \textbf{ID}& \textbf{Issue Type} & \textbf{WCAG 2.1} & \textbf{Platform} & \textbf{Detected Or Repaired By Tool} & \textbf{People Affected}  & \textbf{Source} \\ 
\midrule

\textbf{Operability}& \normalsize{\textcircled{\scriptsize{1}} 
}\normalsize\enspace   & Visual size of touch targets (such as buttons and icons) is too small  & W2.5.5-AAA & app& \checkmark & \begin{tabular}[c]{@{}l@{}}M(MOTOR),\\ BVI(Blind and Visually Impaired)\end{tabular} & \begin{tabular}[c]{@{}l@{}}{\cite{3}},{\cite{51}},{\cite{52}},{\cite{5}},{\cite{15}},{\cite{16}},{\cite{11}},\\ {\cite{53}},{\cite{13}}\end{tabular} \\ 
\cline{2-8}

 {\color[HTML]{000000} }  & \normalsize{\textcircled{\scriptsize{2}} 
}\normalsize\enspace   &  Key UI elements do not maintain a consistent position across screens.  &  -& app &  \checkmark & M  & {\cite{3}},{\cite{51}}\\ 
\cline{2-8}

 {\color[HTML]{000000} }& \normalsize{\textcircled{\scriptsize{3}} 
}\normalsize\enspace  & Pop-up menus or dialogues in the interface cannot be easily closed &  - & app & \checkmark &  M,BVI & {\cite{3}},{\cite{51}} \\
\cline{2-8}
 
{\color[HTML]{000000} }& \normalsize{\textcircled{\scriptsize{4}} 
}\normalsize\enspace   & Icons too close to each other & -   & app  & \checkmark  & M &{\cite{3}},{\cite{51}},{\cite{15} }   \\ 
\cline{2-8}
 
 & \normalsize{\textcircled{\scriptsize{5}} 
}\normalsize\enspace  & Repeated clickable borders & - & app        & \checkmark    & M  & {\cite{52}},{\cite{5}} \\
\cline{2-8}
 
   & \normalsize{\textcircled{\scriptsize{6}} 
}\normalsize\enspace  & Invalid Action  & - & app        & \checkmark   & BVI  & {\cite{28}},{\cite{10}}\\
\cline{2-8}
 
  & \normalsize{\textcircled{\scriptsize{7}} 
}\normalsize\enspace  & Excessive Interaction  & -  & app        & \checkmark  & BVI   &{\cite{28}},{\cite{43}}\\
\cline{2-8}
 
 & \normalsize{\textcircled{\scriptsize{8}} 
}\normalsize\enspace  & Facial recognition doesn't work & - & app,web    & & M   &{\cite{54}}\\
\cline{2-8}
 
& \normalsize{\textcircled{\scriptsize{9}} 
}\normalsize\enspace  & \begin{tabular}[c]{@{}l@{}}Additional head movements in VR manoeuvres by older people \\ interacting with interactive elements outside the field of view\end{tabular} & - & vr& & O(Old)    & {\cite{55}}\\
\cline{2-8}

& \normalsize{\textcircled{\scriptsize{10}} 
}\normalsize\enspace & \begin{tabular}[c]{@{}l@{}}Older adults have difficulty performing interaction tasks in VR \\ that require two-handed coordination\end{tabular}        & - & vr& & O&{\cite{55}}\\
\cline{2-8}

& \normalsize{\textcircled{\scriptsize{11}} 
}\normalsize\enspace & Dialog focus not initialized   & W2.4.7 AA  & web        & \checkmark     & M,BVI,CI（Cognitive impairment） &{\cite{9}}\\
\cline{2-8}

& \normalsize{\textcircled{\scriptsize{12}} 
}\normalsize\enspace & Dialog focus is not limited    & W2.4.7 AA  & web        & \checkmark     & M,BVI,CI  &{\cite{9}}\\
\cline{2-8}

 & \normalsize{\textcircled{\scriptsize{13}} 
}\normalsize\enspace & Dialog box cannot be closed    & W2.1.2 A   & web        & \checkmark     & M,BVI,CI&{\cite{9}}\\
\cline{2-8}

& \normalsize{\textcircled{\scriptsize{14}} 
}\normalsize\enspace & Unable to access certain elements or functions from the keyboard   & W2.1.1 A   & web        & \checkmark     & M,BVI,CI  &{\cite{56}}\\
\cline{2-8}

& \normalsize{\textcircled{\scriptsize{15}} 
}\normalsize\enspace & Keyboard Traps  & W2.1.2 A   & web        & \checkmark     & M,BVI,CI  &{\cite{56}}\\
\cline{2-8}

& \normalsize{\textcircled{\scriptsize{16}} 
}\normalsize\enspace & UI element collision  & - & app        & \checkmark     & BVI       &{\cite{21}}\\
\cline{2-8}

& \normalsize{\textcircled{\scriptsize{17}} 
}\normalsize\enspace & UI elements are missing        & - & app        & \checkmark     & BVI       & {\cite{21}}\\
\cline{2-8}

& \normalsize{\textcircled{\scriptsize{34}} 
}\normalsize\enspace & blank title     & W2.4.6 AA  & web        & & BVI       &{\cite{57}}\\
\cline{2-8}

& \normalsize{\textcircled{\scriptsize{18}} 
}\normalsize\enspace & Top checkbox    & - & web        & & BVI       &{\cite{58}}\\
\cline{2-8}

& \normalsize{\textcircled{\scriptsize{49}} 
}\normalsize\enspace & URLSpan does not use absolute URLs   & W2.4.4 A   & app        & \checkmark     & BVI       & {\cite{52}}\\
\cline{2-8}

& \normalsize{\textcircled{\scriptsize{50}} 
}\normalsize\enspace & \begin{tabular}[c]{@{}l@{}}Screen readers do not navigate elements on the screen \\ according to their hierarchical order\end{tabular}  & W2.4.3 A   & app,web    & \checkmark     & M,BVI     & {\cite{5}},{\cite{10}},{\cite{57}},{\cite{59}}\\
\cline{2-8}

& \normalsize{\textcircled{\scriptsize{51}} 
}\normalsize\enspace & VoiceOver Navigation Loop      & W2.4.3 A   & app        & \checkmark     & BVI       & {\cite{39}},{\cite{60}}\\
\cline{2-8}

& \normalsize{\textcircled{\scriptsize{52}} 
}\normalsize\enspace & ui element cannot be positioned& - & app        & \checkmark     & BVI       &{\cite{39}},{\cite{28}},{\cite{10}}\\
\cline{2-8}

& \normalsize{\textcircled{\scriptsize{53}} 
}\normalsize\enspace & label order     & W2.4.3 A   & web        & & BVI       &{\cite{57}}\\
\cline{2-8}

& \normalsize{\textcircled{\scriptsize{54}} 
}\normalsize\enspace & Layout changes may occur as the app navigates through the screen reader     & - & app        & \checkmark     & BVI       &{\cite{29}},{\cite{43}},{\cite{13}}\\
\cline{2-8}

& \normalsize{\textcircled{\scriptsize{55}} 
}\normalsize\enspace & Neck fatigue in VR use& - & vr& & O& {\cite{55}}\\
\cline{2-8}

& \normalsize{\textcircled{\scriptsize{19}} 
}\normalsize\enspace & Input Method Challenge& - & app        & & O&{\cite{48}}\\
\midrule

\textbf{Perceivability}    & \normalsize{\textcircled{\scriptsize{20}} 
}\normalsize\enspace & Visual details in the video not described in the audio       & W1.2.3 A   & app,web,vr & \checkmark     & BVI       &{\cite{61}},{\cite{62}},{\cite{43}}\\
\cline{2-8}

 & \normalsize{\textcircled{\scriptsize{21}} 
}\normalsize\enspace & Lack of subtitles in the video & W1.2.2 A   & app,web,vr & \checkmark     & DHH(Deaf and Hard of Hearing)  & {\cite{62}}\\
\cline{2-8}

 & \normalsize{\textcircled{\scriptsize{22}} 
}\normalsize\enspace & \begin{tabular}[c]{@{}l@{}}The contrast between the text colour and the background colour \\ is less than 3.0\end{tabular}     & W1.4.3 AA  & app        & \checkmark     & BVI       &{\cite{52}},{\cite{5}},{\cite{64}},{\cite{31}},{\cite{15}},{\cite{11}},{\cite{53}},{\cite{43}}\\
\cline{2-8}

&\normalsize{\textcircled{\scriptsize{23}} 
}\normalsize\enspace & \begin{tabular}[c]{@{}l@{}}The contrast between the foreground and background colours of the image \\ is less than 3.0\end{tabular}     & W1.4.3 AA  & app        & \checkmark     & BVI       & {\cite{52}},{\cite{5}},{\cite{64}},{\cite{31}},{\cite{15}},{\cite{11}},{\cite{53}},{\cite{43}},{\cite{13}}\\
\cline{2-8}

& \normalsize{\textcircled{\scriptsize{24}} 
}\normalsize\enspace & Lack of labelling     & W1.1.1 A   & app,web    & \checkmark     & BVI       &\begin{tabular}[c]{@{}l@{}}{\cite{52}},{\cite{29}},{\cite{11}},{\cite{53}},{\cite{43}},{\cite{13}},{\cite{5}},{\cite{63}},\\ {\cite{22}},{\cite{17}},{\cite{28}},{\cite{38}},{\cite{57}},{\cite{65}},{\cite{66}},{\cite{59}}\end{tabular} \\
\cline{2-8}

& \normalsize{\textcircled{\scriptsize{25}} 
}\normalsize\enspace & Older people's slower spatial perception in VR& - & vr& & O&{\cite{55}}\\
\cline{2-8}

& \normalsize{\textcircled{\scriptsize{26}} 
}\normalsize\enspace & Inappropriate font size        & W1.4.4 AA  & app        & \checkmark     & BVI  &{\cite{53}},{\cite{43}},{\cite{48}},{\cite{18}}\\
\cline{2-8}
 & \normalsize{\textcircled{\scriptsize{27}} 
}\normalsize\enspace & Poor letter spacing   & W1.4.12 AA & app        & & BVI       &  {\cite{53}}\\
\cline{2-8}
& \normalsize{\textcircled{\scriptsize{28}} 
}\normalsize\enspace & Web content is truncated or hidden at small viewport sizes   & W1.4.10 AA & web        & \checkmark     & M,BVI     &{\cite{30}}\\
\cline{2-8}
& \normalsize{\textcircled{\scriptsize{29}} 
}\normalsize\enspace & Button shape issues   & W1.4.1 A   & app        & \checkmark     & BVI       &{\cite{39}}\\
\cline{2-8}
& \normalsize{\textcircled{\scriptsize{30}} 
}\normalsize\enspace & Editable text content & - & app        & \checkmark     & BVI       & {\cite{11}}\\
\cline{2-8}
 & \normalsize{\textcircled{\scriptsize{35}} 
}\normalsize\enspace & text truncation & W1.4.4 AA  & app        & \checkmark     & BVI       &{\cite{21}}\\
\midrule
\textbf{Understandability} & \normalsize{\textcircled{\scriptsize{31}} 
}\normalsize\enspace & Redundant labelling   & - & app        & \checkmark     & CI,BVI    &{\cite{52}},{\cite{5}}\\
\cline{2-8}
 & \normalsize{\textcircled{\scriptsize{32}} 
}\normalsize\enspace & UI elements with the same label on the same screen. & - & app        & \checkmark     & BVI       &{\cite{52}},{\cite{5}}\\
\cline{2-8}
& \normalsize{\textcircled{\scriptsize{33}} 
}\normalsize\enspace & Incorrect label content for UI elements       & - & web        & \checkmark     & BVI       & {\cite{38}}\\
\cline{2-8}
& \normalsize{\textcircled{\scriptsize{36}} 
}\normalsize\enspace & Web page cannot translate content into sign language& - & web        & & DHH       & {\cite{67}}\\
\cline{2-8}
& \normalsize{\textcircled{\scriptsize{37}} 
}\normalsize\enspace & Complex navigation    & - & app        & & O& {\cite{48}}\\
\cline{2-8}
 & \normalsize{\textcircled{\scriptsize{38}} 
}\normalsize\enspace & Unintuitive design of the interface  & - & app        & & O&{\cite{48}}\\
\cline{2-8}
 & \normalsize{\textcircled{\scriptsize{39}} 
}\normalsize\enspace & Rapid dialogue or slurred pronunciation       & - & app,web,vr & & A(aphasia)&{\cite{68}}\\
\cline{2-8}
 & \normalsize{\textcircled{\scriptsize{40}} 
}\normalsize\enspace & The fast pace of audiovisual media   & - & app,web,vr & & A&{\cite{68}}\\
\cline{2-8}
& \normalsize{\textcircled{\scriptsize{41}} 
}\normalsize\enspace & Insufficient reading time for on-screen text  & - & app,web,vr & & A&{\cite{68}}\\
\cline{2-8}
& \normalsize{\textcircled{\scriptsize{42}} 
}\normalsize\enspace & Complex narrative structure    & - & app,web,vr & & A& {\cite{68}}\\
\cline{2-8}
& \normalsize{\textcircled{\scriptsize{43}} 
}\normalsize\enspace & Redundant information & - & app        & \checkmark     & BVI       &{\cite{18}},{\cite{69}}\\
\cline{2-8}
 & \normalsize{\textcircled{\scriptsize{44}} 
}\normalsize\enspace & EditText and editable TextView with non-empty contentDescription   & - & app        & \checkmark     & BVI       &{\cite{52}}\\
\midrule
\textbf{Robustness}& \normalsize{\textcircled{\scriptsize{45}} 
}\normalsize\enspace & ui elements are not supported by text scaling & W4.1.2 A   & app        & \checkmark     & BVI       &{\cite{4}},{\cite{39}}\\
\cline{2-8}
& \normalsize{\textcircled{\scriptsize{46}} 
}\normalsize\enspace & ui elements are not supported by screen readers     & W4.1.2 A   & app,web    & \checkmark     & BVI       &\begin{tabular}[c]{@{}l@{}}{\cite{52}},{\cite{29}},{\cite{60}},{\cite{43}},{\cite{13}},{\cite{5}},{\cite{64}},{\cite{10}},\\ {\cite{59}}\end{tabular}\\
\cline{2-8}
& \normalsize{\textcircled{\scriptsize{47}} 
}\normalsize\enspace & Using AJAX and JavaScript      & W4.1.2 A   & web        & & BVI       &{\cite{70}}\\
\cline{2-8}
& \normalsize{\textcircled{\scriptsize{48}} 
}\normalsize\enspace & compatibility issue   & - & app        & & O&{\cite{48}}\\
\bottomrule
\end{tabular}}
\vspace{-1.5em}
\end{table*}

The survey was launched on August 27, 2024, and ended on September 18, 2024, with a total of 130 responses collected. Among the 130 respondents, end-users represented the largest group (55.38\%), followed by academic researchers (11.54\%), other occupations (12.31\%), developers (10.77\%),testers (6.15\%), and designers (3.85\%). Most participants were aged between 18 and 30 years, with 40.77\% aged 18-25, 33.08\% aged 26-30, 16.92\% aged 31-40. Participants under 18 made up 5.38\%, while those over 41 accounted for just 4.62\%.

The survey was conducted using the Wjx.cn platform for questionnaire design, data collection, and analysis. This platform is widely used for survey design and data management. 

\subsection{Analysis of Tool Capabilities}
\label{section:Analysis of Tool Capabilities}
We conducted an in-depth analysis of the 42 selected articles (refer to Section \ref{section:Systematic Literature Review}) with the aim of identifying all automated tools capable of detecting and repairing accessibility issue types, and further analyzing these tools. This process encompassed three steps: tool selection, capability information collection, and analysis based on the taxonomy, as illustrated in the third module of Figure \ref{Framework}. Importantly, we did not execute these tools ourselves; rather, we relied on the reported capabilities from their performance on the respective datasets.

Firstly, in the tool selection step, to ensure targeted analysis and simplify the selection process, we established the following criteria, and only tools meeting these criteria were included in the analysis:
\begin{itemize}
\item \textbf{\emph{Criterion \#1}}:Tools must be directly related to the detection or repair of accessibility issues, with descriptions containing keywords or synonyms like “identification,” “detection,” “repair,” and “accessibility issues.”
\item \textbf{\emph{Criterion \#2}}:Tools must be automated, able to perform detection or repair tasks independently without manual intervention.
\item \textbf{\emph{Criterion \#3}}:Tool functionality must be clearly described, including specific operating principles, technical implementation, and the process of identifying or repairing accessibility issues.
\item \textbf{\emph{Criterion \#4}}:Tools must explicitly state the accessibility issues they can detect or repair, with related documentation or descriptions listing supported issues and explaining their capabilities.
\end{itemize}

Based on these criteria, we collected 14 detection tools and 9 repair tools from 42 related papers.

We then collected information on tool capabilities. In this step, we first identified key information for screening, including creation time, name, detectable or repairable accessibility issues, main techniques used, training set, and validation set, among others. We then thoroughly reviewed the papers proposing these tools, extracting the key information from sections like “Methods” and “Results” and compiling it into a table.In instances where tools lacked a designated name, we adopted an abbreviation strategy, leveraging the initial letters of the first three authors of the corresponding paper.

Finally, we conducted a detailed analysis of tool capabilities based on the taxonomy we developed, covering tool performance in detecting and repairing issue types within the taxonomy, application scenarios, and evaluation metrics. First, we evaluated the number of issue types each tool could detect or repair based on the constructed taxonomy.  For this, we introduced the metric “ratio of successfully detected or repaired types” to measure the tools’ capabilities, calculated as: the number of successfully detected or repaired issue types divided by the total number of issue types. Next, we analyzed the application scenarios of each tool to understand their suitability for different platforms (e.g., mobile applications, website). Finally, we examined the evaluation metrics used by each tool, focusing on metrics like precision and recall.

\begin{table*}[htbp]
\vspace{-1.5em}
\caption{Tools List for Accessibility Issues Detection.}
\label{detect tools}
\centering
\resizebox{1.0\linewidth}{!}{
\begin{tabular}{lllllllllllll}
\toprule

\textbf{Detection Issues ID} & \textbf{Platform}   & \textbf{Year} & \textbf{Name} & \textbf{Technology}  & \textbf{\begin{tabular}[c]{@{}l@{}}Inaccessible \\ Rate（\%）\end{tabular}} & \textbf{\begin{tabular}[c]{@{}l@{}}Activity \\ Coverage(\%)\end{tabular} }& \textbf{Accuracy(\%)}  & \textbf{Precision(\%)} & \textbf{F1-Score} & \textbf{Recall(\%)} & \textbf{Sources}  & \textbf{Useable} \\

\midrule
\normalsize{\textcircled{\scriptsize{1}} \textcircled{\scriptsize{2}} \textcircled{\scriptsize{3}} \textcircled{\scriptsize{4}}
}\normalsize\enspace     & Mobile     & 2024 & {\color[HTML]{060607} MotorEase} & Faster-RCNN,Optical Character Recognition        & -    & -    & \begin{tabular}[c]{@{}l@{}}85.25,87.76,\\ 91.23,95.75\end{tabular} & \begin{tabular}[c]{@{}l@{}}100,82.14,\\ 90.42,71.19\end{tabular} & \begin{tabular}[c]{@{}l@{}}79.86,88.46,\\ 91.29,83.17\end{tabular} & - & {\cite{3}} & 1       \\
\midrule
\normalsize{\textcircled{\scriptsize{28}} 
}\normalsize\enspace   & Web        & 2024 & SALAD    & Natural Language Processing,Mechine Learning     & -    & -    & -      & 85   & -      & 94& {\cite{30}} & 0       \\
\midrule
\normalsize{\textcircled{\scriptsize{6}} \textcircled{\scriptsize{7}} \textcircled{\scriptsize{24}} \textcircled{\scriptsize{52}}
}\normalsize\enspace    & Mobile     & 2024 & AdMole   & Action Extractor,Snapshot Manager & 84.4 & -    & -      & -    & -      & - & {\cite{28}} & 1       \\
\midrule
\normalsize{\textcircled{\scriptsize{29}} \textcircled{\scriptsize{45}} \textcircled{\scriptsize{51}} \textcircled{\scriptsize{52}}
}\normalsize\enspace  & Mobile     & 2024 & AXNAV    & Large Language Model     & -    & -    & 70     & -    & -      & - & {\cite{39}}  & 0       \\
\midrule
\normalsize{\textcircled{\scriptsize{11}} \textcircled{\scriptsize{12}} \textcircled{\scriptsize{33}}
}\normalsize\enspace     & Web        & 2023 & LOTUS    & Keyboard Dialog Flow Graph,Stacking Model        & -    & -    & -      & 71,70,40   & -      & 83,90,83   & {\cite{9}} & 0       \\
\midrule
\normalsize{\textcircled{\scriptsize{45}} 
}\normalsize\enspace   & Mobile     & 2022 & AccessiText       & Optical Character Recognition,Heuristic Rule     & -    & -    & -      & 87.59& -      & 95.3       & {\cite{4}} & 0       \\
\midrule
\normalsize{\textcircled{\scriptsize{20}} \textcircled{\scriptsize{21}} 
}\normalsize\enspace      & Mobile,Web & 2022 & CrossA11y& Multimodal Machine Learning,Cross-modal Grounding& -    & -    & -      & 69.4,98.3  & 81.4,90.8    & 98.4,84.3  & {\cite{62}} & 1       \\
\midrule
\begin{tabular}[c]{@{}l@{}}
\normalsize{\textcircled{\scriptsize{1}} \textcircled{\scriptsize{5}} \textcircled{\scriptsize{22}} \textcircled{\scriptsize{23}} \textcircled{\scriptsize{24}}
}\normalsize\enspace \\ \normalsize{\textcircled{\scriptsize{31}} \textcircled{\scriptsize{32}} \textcircled{\scriptsize{44}} \textcircled{\scriptsize{46}} \textcircled{\scriptsize{49}}
}\normalsize\enspace \end{tabular} & Mobile     & 2022 & Xbot     & Instrumentation Technique),Static Program Analysis)    & -    & 72.81& -      & -    & -      & - & {\cite{52}} & 1       \\
\midrule
\normalsize{\textcircled{\scriptsize{46}} \textcircled{\scriptsize{51}} 
}\normalsize\enspace    & Mobile     & 2022 & ATARI    & Graph-based Models,Dynamic Analysis     & -    & -    & -      & 92   & -      & 94& {\cite{60}} & 0       \\
\midrule
\normalsize{\textcircled{\scriptsize{6}} \textcircled{\scriptsize{46}} \textcircled{\scriptsize{52}} 
}\normalsize\enspace   & Mobile     & 2022 & Groundhog& Action Extractor,Snapshot Manager & -    & 98   & -      & 86   & -      & 83& {\cite{10}} & 1       \\
\midrule
\normalsize{\textcircled{\scriptsize{1}} \textcircled{\scriptsize{24}} \textcircled{\scriptsize{46}} \textcircled{\scriptsize{54}} 
}\normalsize\enspace  & Mobile     & 2021 & Latte    & Dynamic Program Analysis & -    & -    & -      & 100  & -      & - & {\cite{13}} & 1       \\
\midrule
\normalsize{\textcircled{\scriptsize{14}} \textcircled{\scriptsize{15}} 
}\normalsize\enspace    & Web        & 2021 & {\color[HTML]{060607} KAFE}      & \begin{tabular}[c]{@{}l@{}}Keyboard Navigation Flow Graph,\\ Point-Click Navigation Flow Graphs\end{tabular} & -    & -    & -      & 92,90& -      & 100,100    & {\cite{56}} & 0       \\
\midrule
\normalsize{\textcircled{\scriptsize{24}} \textcircled{\scriptsize{33}} 
}\normalsize\enspace    & Web        & 2021 & AXERAY   & \begin{tabular}[c]{@{}l@{}}Convolutional Neural Network,\\ Natural Language Processing\end{tabular} & -    & -    & 85     & -    & 87     & - & {\cite{38}} & 1       \\
\midrule
\normalsize{\textcircled{\scriptsize{1}} \textcircled{\scriptsize{23}} \textcircled{\scriptsize{24}} \textcircled{\scriptsize{30}} 
}\normalsize\enspace & Web        & 2017 & AIFA     & Axe-Android        & \begin{tabular}[c]{@{}l@{}}59.4,30.27,\\ 46.34,1.33\end{tabular} & -    & -      & -    & -      & - & {\cite{11}} & 1\\

\bottomrule
\end{tabular}}

\end{table*}

\begin{table*}

\caption{Tools List for Accessibility Issues Repair.}
\label{repair tools}
\centering
\resizebox{1.0\linewidth}{!}{
\begin{tabular}{lllllllllllll}
\toprule
\textbf{Repair Issues ID} & \textbf{Platform} & \textbf{Year} & \textbf{Name}      & \textbf{Technology}    & \textbf{Accuracy(\%)}       & \textbf{Recall(\%)} & \textbf{F1-Score（\%）} & \textbf{Exact match（\%）} & \textbf{\begin{tabular}[c]{@{}l@{}}Repair Success\\ Rate（\%）\end{tabular}} & \textbf{\begin{tabular}[c]{@{}l@{}}Accessibility\\ Rate（\%）\end{tabular}} & \textbf{Sources} & \textbf{Usable} \\
\midrule
\normalsize{\textcircled{\scriptsize{1}} \textcircled{\scriptsize{4}} \textcircled{\scriptsize{22}} \textcircled{\scriptsize{23}} 
}\normalsize\enspace & Mobile   & 2024 & AccessFixer        & \begin{tabular}[c]{@{}l@{}}Relational-Graph, \\ Convolutional Neural Network\end{tabular}   & -   & - & -   & -& 83.75,80.7,80.3,80.3 & -    & {\cite{15}} & 1      \\
\midrule
\normalsize{\textcircled{\scriptsize{22}} \textcircled{\scriptsize{23}} 
}\normalsize\enspace     & Mobile   & 2023 & Iris& Context-aware Color Selection   & -   & - & -   & -& 93.6,70.4   & -    & {\cite{31}} & 1      \\
\midrule
\normalsize{\textcircled{\scriptsize{26}} \textcircled{\scriptsize{43}} 
}\normalsize\enspace     & Mobile   & 2023 & AGAA& \begin{tabular}[c]{@{}l@{}}Genetic Algorithm,\\ Density-Based Clustering Algorithm\end{tabular}   & -   & - & -   & -& 95.3,83.8   & -    & {\cite{18}}  & 1      \\
\midrule
\normalsize{\textcircled{\scriptsize{24}} 
}\normalsize\enspace       & Web      & 2023 & SAM & Regularized Method     & 93.4& 91.9       & 92.6& -& -     & -    & {\cite{20}} & 1      \\
\midrule
\normalsize{\textcircled{\scriptsize{16}} \textcircled{\scriptsize{17}} \textcircled{\scriptsize{35}}
}\normalsize\enspace  & Mobile   & 2023 & ScaleFix  & \begin{tabular}[c]{@{}l@{}}Optical Character Recognition,\\ Multi-Objective Genetic Algorithm\end{tabular} & -   & - & -   & -& 90    & -    & {\cite{21}} & 0      \\
\midrule
\normalsize{\textcircled{\scriptsize{1}}
}\normalsize\enspace& Mobile   & 2021 & SALEM     & Size Relation Graph    & -   & - & -   & -& 99    & 99   & {\cite{16}} & 1      \\
\midrule
\normalsize{\textcircled{\scriptsize{24}} 
}\normalsize\enspace      & Mobile   & 2021 & coala     & \begin{tabular}[c]{@{}l@{}}Recurrent Neural Network,\\ Convolutional Neural Network\end{tabular}  & -   & - & -   & 38     &       & -    & {\cite{17}} & 1      \\
\midrule
\normalsize{\textcircled{\scriptsize{24}} \textcircled{\scriptsize{46}} \textcircled{\scriptsize{50}} 
}\normalsize\enspace & Mobile   & 2021 & Screen Recognition & \begin{tabular}[c]{@{}l@{}}Optical Character Recognition,\\ Heuristic Rule\end{tabular}     & {\color[HTML]{060607} 71.3} & - & -   & -& -     & -    & {\cite{22}} & 0      \\
\midrule
\normalsize{\textcircled{\scriptsize{24}} 
}\normalsize\enspace        & Mobile   & 2020 & LabelDroid& \begin{tabular}[c]{@{}l@{}}Convolutional Neural Network,\\ Transformer Model\end{tabular}   & -   & - & -   & 60.7   & -     & -    & {\cite{19}} & 1     
 \\
\bottomrule

\end{tabular}}
\vspace{-1.5em}
\end{table*}

\subsection{Analysis of Dataset Status}
\label{section:Analysis of Dataset Status}
In order to comprehensively understand the current status of datasets used for accessibility issue detection and repair tools, we conducted a filtering and analysis of relevant datasets. This process comprised three steps: dataset selection, key information collection, and taxonomy-based analysis, as illustrated in the fourth module of Figure \ref{Framework}.

Firstly, in terms of dataset selection, to ensure targeted analysis and dataset usability, we established the following criteria, and only datasets meeting these criteria were included in the subsequent analysis:

\begin{itemize}
\item \textbf{\emph{Criterion \#1}}:Datasets should explicitly contain screenshots or videos of webpages or mobile applications with accessibility issues, providing direct instances of accessibility issues for research.
\item \textbf{\emph{Criterion \#2}}:The dataset must be publicly available and accessible without restrictions, as disclosed by its authors.
\end{itemize}

Based on these criteria, we identified 10 datasets for detection tools and 8 for repair tools from the 42 selected articles.

Secondly, we collected key information on these datasets. Specifically, we identified the essential information needed for systematic documentation of each dataset, including dataset name, data type (e.g., Apk, Image), the number of instances in the dataset, types of accessibility issues covered, and dataset creation time, among other details. We then thoroughly reviewed the relevant papers, extracting this information from sections like “Introduction,” “Methods,” and “Results,” and compiled it into a table.

Finally, We systematically analyzed the current status of these datasets across three aspects: coverage scope, data types, and scale. To assess the coverage scope of these datasets, we examined the alignment of each dataset with the constructed taxonomy of accessibility issue types. For this, we introduced a “coverage rate” metric to measure the coverage scope of these datasets, calculated as: the number of issue types included in the dataset divided by the total number of issue types. We then analyzed the data type of the datasets and evaluated the dataset scale by the number of instances.

\section{Results and Findings}
Building on the framework outlined in section \ref{section:Framework}, this section provides responses to the three research questions.
\subsection{Taxonomy}
For RQ1, we constructed a comprehensive taxonomy of accessibility issues types following the process outlined in Section \ref{section:Construction of the Taxonomy}  

This taxonomy encompasses 55 types of accessibility issues, systematically categorized into four main groups: Operability, Perceivability, Understandability, and Robustness, as shown in Table \ref{Taxonomy}.

We present the taxonomy analysis results from three distinct perspectives: the categories, the comprehensive analysis of the taxonomy, and the result of the questionnaire survey.

\subsubsection{Categories}
Rregarding specific categories, the first category, ``Operability," which requires that the user interface and navigation be operable by all users, includes 27 accessibility issue types, accounting for 49.09\% of the total and representing the category with the most issue types. These issue types primarily affect users with motor and visual impairments, highlighting the significant challenges they face in interface interaction and operation. Among these 27 issue types, 11 have explicit WCAG 2.1 guidance, and 19 can be detected or repaired by existing tools.
The second category, ``Perceivability," emphasizing that information and user interface components must be perceivable, contains 12 issue types, mainly affecting users with visual and auditory impairments. Of these 12 issue types, 10 have explicit WCAG 2.1 guidance, and 10 can be detected or repaired by tools.
The third category, ``Understandability," requiring that information and interfaces be clear and straightforward, includes 12 issue types. These issue types primarily affect users with visual impairments and aphasia, centering around information comprehension challenges. None of these issue types have explicit WCAG 2.1 guidance, while five can be identified and addressed by detection and repair tools.
The fourth category, ``Robustness," which ensures a consistent and accessible experience regardless of technology or tools used, encompasses 4 issue types, mainly affecting users with visual impairments and older adults. Among these 4 issue types, 3 have WCAG guidance, and 2 can be detected or repaired by existing tools. 

In comparison, the operability category encompasses the highest number of issue types, suggesting that users face greater obstacles during interaction. Moreover, none of the issue types in the understandability category is explicitly covered by current guidelines, underscoring a limitation of WCAG 2.1 in addressing understandability-related issues.

\begin{table*}
\caption{Datasets List for Accessibility Issues Detection and Repair.}
\label{Datasets}
\centering
\renewcommand{\arraystretch}{1.2}
\resizebox{1.0\linewidth}{!}{
\begin{tabular}{lllllll}
\toprule
\textbf{Use} & \textbf{Name} & \textbf{Data Type} & \textbf{Instances} & \textbf{Covery Issues ID} & \textbf{Used by}  & \textbf{Year} \\

\midrule
\textbf{Detection} & TSO    & Apk & 248 & \normalsize{\textcircled{\scriptsize{1}} \textcircled{\scriptsize{23}} \textcircled{\scriptsize{24}} \textcircled{\scriptsize{30}} 
}\normalsize\enspace  & {\cite{11}} & 2023 \\
\cline{2-7}
     & CCF    & Apk & 2,270     & \begin{tabular}[c]{@{}l@{}}\normalsize{\textcircled{\scriptsize{1}} \textcircled{\scriptsize{5}} \textcircled{\scriptsize{22}} \textcircled{\scriptsize{23}} \textcircled{\scriptsize{24}} 
}\normalsize\enspace \\ \normalsize{\textcircled{\scriptsize{31}} \textcircled{\scriptsize{32}} \textcircled{\scriptsize{44}} \textcircled{\scriptsize{46}} \textcircled{\scriptsize{49}} 
}\normalsize\enspace
   \end{tabular} & {\cite{52}} & 2022 \\\cline{2-7}
     & SMM    & Apk & 57  & \normalsize{\textcircled{\scriptsize{1}} \textcircled{\scriptsize{5}} \textcircled{\scriptsize{22}} \textcircled{\scriptsize{23}} \textcircled{\scriptsize{24}} 
}\normalsize\enspace    & {\cite{10}} & 2022 \\\cline{2-7}
     & SAL    & Apk & 20  & \normalsize{\textcircled{\scriptsize{1}} \textcircled{\scriptsize{24}} \textcircled{\scriptsize{46}} \textcircled{\scriptsize{54}} 
}\normalsize\enspace & {\cite{13}} & 2021 \\\cline{2-7}
     & HHM    & Images    & 500 & \normalsize{\textcircled{\scriptsize{6}} \textcircled{\scriptsize{7}} \textcircled{\scriptsize{24}} \textcircled{\scriptsize{52}}
}\normalsize\enspace & {\cite{28}} & 2024 \\\cline{2-7}
     & MotorCheck   & Images,Xml& 1,599     & \normalsize{\textcircled{\scriptsize{1}} \textcircled{\scriptsize{2}} \textcircled{\scriptsize{3}} \textcircled{\scriptsize{4}} 
}\normalsize\enspace    & {\cite{3}}{\cite{51}} & 2024 \\\cline{2-7}
     & YouDescribe  & Video     & 20  & \normalsize{\textcircled{\scriptsize{20}} \textcircled{\scriptsize{21}} 
}\normalsize\enspace     & {\cite{62}} & 2022 \\\cline{2-7}
     & AudioSet     & Video     & 2,000,000 & \normalsize{\textcircled{\scriptsize{20}} \textcircled{\scriptsize{21}} 
}\normalsize\enspace           & {\cite{62}} & 2017 \\\cline{2-7}
     & HowTo100M    & Video     & 1,000,000 & \normalsize{\textcircled{\scriptsize{20}} \textcircled{\scriptsize{21}} 
}\normalsize\enspace           & {\cite{62}} & 2019 \\\cline{2-7}
     & BM     & Web & 30  & \normalsize{\textcircled{\scriptsize{24}} \textcircled{\scriptsize{33}} 
}\normalsize\enspace           & {\cite{38}} & 2021 \\

\midrule
\textbf{Repair}     & ZCF    & Apk & 10,078    & \normalsize{\textcircled{\scriptsize{22}} \textcircled{\scriptsize{23}} 
}\normalsize\enspace        & {\cite{31}} & 2023 \\\cline{2-7}
     & ACH    & Apk & 48  & \normalsize{\textcircled{\scriptsize{1}} 
}\normalsize\enspace      & {\cite{16}} & 2021 \\\cline{2-7}
     & XLL    & Images    & 31  & \normalsize{\textcircled{\scriptsize{26}} \textcircled{\scriptsize{43}} }\normalsize\enspace           & {\cite{18}}  & 2023 \\\cline{2-7}
     & ZZG-1  & Images    & 18,317    & \normalsize{\textcircled{\scriptsize{24}} }\normalsize\enspace     & {\cite{20}} & 2023 \\\cline{2-7}
     & LabelDroid-ex& Images    & 21,864    & \normalsize{\textcircled{\scriptsize{24}} }\normalsize\enspace& {\cite{17}} & 2021 \\\cline{2-7}
     & LabelDroid   & Images    & 19,233    & \normalsize{\textcircled{\scriptsize{24}} }\normalsize\enspace& {\cite{19}} & 2020 \\\cline{2-7}
     & ZLC    & Images,Xml& 2,209     & \normalsize{\textcircled{\scriptsize{1}} \textcircled{\scriptsize{4}} \textcircled{\scriptsize{22}} \textcircled{\scriptsize{23}} }\normalsize\enspace  & {\cite{15}} & 2024 \\\cline{2-7}
     & ZZG-2  & Web & 30  & \normalsize{\textcircled{\scriptsize{24}} }\normalsize\enspace& {\cite{20}} & 2023\\
\bottomrule

\end{tabular}}
\end{table*}

\subsubsection{Comprehensive Analysis: WCAG 2.1 Coverage, Platforms, and Affected User Groups}
In our comprehensive analysis of the taxonomy, we evaluated the coverage of WCAG 2.1, identified the user groups affected, and examined the platforms where these issue types arise. Specifically, in terms of guideline coverage, out of the 55 accessibility issue types, only 24 have explicit WCAG 2.1 guidance, accounting for 43.6\%. The remaining 31 issue types are not adequately covered in WCAG 2.1, highlighting the limitations of the current accessibility guidelines and the need for future research and expansion.
For primary affected user groups,  Among the 55 issue types, 35 impact visually impaired users (63.6\%). In comparison, users with motor impairments face 13 issue types, those with auditory impairments encounter 2 issue types, aphasia patients are affected by 4 issue types, users with cognitive impairments face 6 issue types, and older adults encounter 8 issue types. Visually impaired users face particularly significant challenges, warranting priority consideration in design and development.
Regarding the platforms where these issue types arise, out of the 55 issue types, 29 occur in mobile applications (52.7\%), 12 in websites, 4 in both mobile applications and websites , 4 in VR applications, and 6 in all application contexts. This suggests that mobile applications are the primary context for accessibility issue types, underscoring the need for developers to focus on enhancing mobile accessibility.

\begin{table*}
\caption{Feedback Results of Questionnaire Survey.}
\label{Questionnaire Survey}
\centering
\resizebox{1.0\linewidth}{!}{
\begin{tabular}{ccccccc}
\toprule
\textbf{\#Feedback} &\textbf{ \#Valid Feedback} &\textbf{Rationality} & \textbf{Completeness} &\textbf{Understandability} & \textbf{\begin{tabular}[c]{@{}l@{}}People Affected\\Annotation\end{tabular} }& \textbf{\begin{tabular}[c]{@{}l@{}}WCAG 2.1\\Annotation\end{tabular}}  \\ 
\midrule
130        & 128     & 86.92\%     & 86.16\%      & 66.15\%  & 84.61\% & 85.39\% \\
\bottomrule
\end{tabular}}
\end{table*}

\subsubsection{Questionnaire Survey Result}
Regarding the result of the questionnaire survey, our taxonomy received broad endorsement from various sectors. As shown in Table \ref{Questionnaire Survey}, over 86\% of the 128 valid respondents rated the taxonomy’s rationality and completeness as good or very good, with scores of 4 or 5. More than 60\% provided positive feedback on the taxonomy’s understandability. Over 80\% believed that our annotations for the people affected and WCAG 2.1 were appropriate. Additionally, 10 respondents suggested providing screenshots or videos for each issue type to enhance understanding, which explains the slightly lower comprehensibility score. Additionally, 41.54\% of respondents were aware of accessibility issue types in their applications, 7.69\% were unaware, and 50.77\% were uncertain, reflecting a general lack of awareness regarding accessibility issues. 

Overall, the survey results indicate widespread recognition of our taxonomy’s rationality , completeness, and annotation accuracy, though there remains room for improvement in comprehensibility.

\begin{tcolorbox}[colback=gray!10,
			colframe=black,
			width=9cm,
			arc=2mm, auto outer arc]		
			\textbf{Answer to RQ1}:Through a review of 42 relevant literature, we constructed a taxonomy encompassing 55 accessibility issue types, grouped into four major categories: Operability, Perceivability, Understandability, and Robustness. This taxonomy has received broad endorsement from various sectors, with over 84\% of feedback indicating high ratings for reasonableness, completeness, and accuracy in annotating people affected and WCAG 2.1.
\end{tcolorbox}

\subsection{Capabilities of Detection and Repair Tools}
To address RQ2, this section aims to assess the capabilities of existing detection and repair tools collected from the literature, based on the constructed taxonomy of accessibility issue types. Following the methodology outlined in Section \ref{section:Analysis of Tool Capabilities}, we identified 14 representative detection tools and 9 repair tools, as detailed in Table \ref{detect tools} and Table \ref{repair tools}.

Next, we present the analysis results from two main perspectives: detection tools and repair tools.

\subsubsection{Capabilities of Detection Tools}
We presents the analysis results for 14 detection tools, including MotorEase, SALAD, AccessiText, CrossA11y, and Xbot,among others, focusing on detection capability, application platforms, and evaluation metrics.

Firstly, regarding the detection capabilities of these tools, these 14 tools collectively identify 31 out of the 55 accessibility issue types in the taxonomy, achieving a detection success rate of 56.3\% (i.e., 31/55). Among these, the majority of issue types can be detected by one or two tools. Notably, only one issue type (i.e., 'Lack of Label') can be detected by multiple tools, including Xbot, AIFA, AdMole, Latte, and AXERAY. Interestingly, none of these 31 issue types are detectable by every tool. Considering the capabilities of individual tools, their detection ranges vary from 1 to 10 issue types. Specifically, Xbot exhibits the highest detection capability, identifying up to 10 issue types, whereas SALAD and AccessiText can each detect only one issue type, with the remaining tools capable of detecting between 2 to 5 issue types. 
Overall, compared to the 55 accessibility issue types we have cataloged, the current detection tools exhibit a relatively low detection rate, with the highest success rate at 18.1\%. This underscores the inadequacy of the detection capabilities of existing tools.

Secondly, concerning application platforms, our findings reveal a prevalence of tools tailored for mobile applications over those designed for websites. Specifically, tools such as MotorEase, AccessiText, Xbot, ATARI, Groundhog, AIFA, AdMole, Latte, and AXNAV are dedicated to mobile application issue detection, whereas SALAD, LOTUS, KAFE, and AXERAY are focused on web application issues. Regrettably, none of these tools offer dual support for both mobile applications and websites , which points to a lack of cross-platform detection capabilities.

Finally, regarding evaluation metrics, the detection tools employ a diverse set of metrics. For instance, SALAD, AccessiText, LOTUS, ATARI, and KAFE prioritize precision and recall, while Xbot, Groundhog, AIFA, and AdMole concentrate on activity coverage and inaccessibility rate.The remaining tools also consider accuracy and the F1 score. This diversity in metrics hinders straightforward performance comparisons. For example, although MotorEase, Xbot, AIFA, and Latte all detect the "Touch Target Size" issue type, each employs a unique metric: MotorEase utilizes accuracy, precision, and the F1 score, Xbot focuses on activity coverage, AIFA measures inaccessibility rate, and Latte considers precision alone. Such diversity complicates the direct identification of the most effective tool for a particular issue type.

In summary, the detection tools collectively address only 31 out of 55 accessibility issue types in the taxonomy, resulting in a detection success rate of 56.3\%. Furthermore, 24 issue types remain undetectable, highlighting substantial room for improvement in detection capabilities. In terms of application contexts, tools designed for mobile applications are more prevalent than those for websites, and the diverse evaluation metrics among tools complicate direct performance comparisons.

\subsubsection{Capabilities of Repair Tools}
We details the analysis results for 9 repair tools, including Iris, AccessFixer, SALEM, and ScaleFix (among others), with respect to repair capability, application platforms, and evaluation metrics.

Firstly, in terms of repair capability, the 9 repair tools collectively address 13 out of the 55 issue types in the taxonomy, yielding a repair success rate of 23.6\%. Notably, the "Lack of Label" issue type is the only one that can be repaired by multiple tools, including coala, LabelDroid, SAM, and Screen Recognition. However, none of these 13 issue types are universally repairable by all tools, further emphasizing the limited capabilities of current repair tools. 
The number of issue types each tool can repair ranges from 1 to 4. Specifically, AccessFixer exhibits the highest repair capability, addressing 4 issue types, whereas SALEM, coala, LabelDroid, and SAM can each repair only one issue type. The remaining tools can repair between 2 and 3 issue types. Overall, the number of issue types that repair tools can address is significantly lower than the number detectable by detection tools. Compared to the 55 issue types identified in the taxonomy, existing repair tools show a much lower capacity for repair, with the highest success rate being a mere 7.2\%. This finding suggests that future research should aim to expand the scope of repair tools to address a wider range of issues types.

Secondly, regarding application platforms, we observe a significant predominance of repair tools for mobile applications over those for websites. Specifically, eight tools—namely Iris, AccessFixer, SALEM, coala, AGAA, LabelDroid, ScaleFix, and Screen Recognition—are tailored for addressing accessibility issue types within mobile applications, while SAM is the sole tool dedicated to websites. Like detection tools, there are no repair tools that currently offer dual support for both mobile applications and websites. In summary, the preponderance of repair tools is directed at mobile applications, with a conspicuous lack of tools catering to both mobile applications and websites, underscoring the limitations of cross-platform repair capabilities.

Finally,in the realm of evaluation metrics,repair tools employ a variety of evaluation metrics. For instance, tools such as Iris, AccessFixer, AGAA, and ScaleFix prioritize the success rate of repairs, while coala and LabelDroid utilize the exact match rate as their primary criterion. Other tools factor in metrics such as accuracy,the recall, the F1 score, and the accessibility rate. As with detection tools, the heterogeneity of evaluation metrics hinders straightforward performance comparisons among tools, making the selection of the most appropriate tool a more intricate process.

In summary, repair tools address a more limited range of issue types compared to detection tools, tackling just 13 out of the 55 issues in the taxonomy, and achieving a modest repair success rate of only 23.6\%. Furthermore, a mere one repair tool is designed for websites, underscoring the paucity of tools dedicated to website repairs, while the diverse set of evaluation metrics further complicates direct performance assessments.

\begin{tcolorbox}[colback=gray!10,
			colframe=black,
			width=9cm,
			arc=2mm, auto outer arc]		
			\textbf{Answer to RQ2}: Collectively, the 14 detection tools identify 31 of the 55 issue types in the taxonomy, achieving a detection success rate of 56.3\%. The 9 repair tools address only 13 of these issues, with a repair success rate of 23.6\%. The diversity in evaluation metrics across tools complicates direct performance comparisons, posing challenges in selecting the most effective tools.
\end{tcolorbox}

\subsection{Dataset Status}
To address RQ3, our aim is to evaluate the current state of pertinent datasets based on  the established taxonomy. In accordance with the methodology outlined in Section \ref{section:Analysis of Dataset Status}, we identified 18 publicly accessible datasets. Among these, 10 datasets are used for detection tools, while the remaining 8 are used for repair tools, as shown in Table \ref{Datasets},

We present the analysis results from two aspects: datasets for detection tools and datasets for repair tools.

\subsubsection{Status of Datasets for Detection Tools}
We offers an in-depth examination of the 10 datasets employed by detection tools, encompassing datasets such as TSO, CCF, SMM, SAL, among others, with a focus on the coverage of issue types, data types, and the quantities of instances.

Firstly, regarding the coverage of issue types, these 10 datasets collectively cover 21 of the 55 accessibility issue types in our taxonomy, achieving an overall coverage rate of 38.1\%. However, when considered individually, each dataset’s coverage range narrows to between 2 and 10 issue types. Among these, the CCF dataset covers the most issue types (10 types), whereas YouDescribe, AudioSet, HowTo100M, and BM datasets each cover only 2 types, showing significant variation across datasets. Notably, issue type 24 (“Lack of Labels”) appears most frequently across TSO, CCF, SAL, HHM, and BM datasets and is found in both mobile applications and websites. This finding indicates the prevalence of this issue type in accessibility detection. In contrast, other types such as “Editable Text Content” (type 30) appear only in the TSO dataset, while “UI Element Unlocatable” (type 52) appears only in SMM, and “Incorrect Label Content” (type 33) only in BM. These infrequent issue types, appearing in just one dataset each, suggest lower prominence across current datasets. Overall, these datasets cover only 21 of the 55 types in the taxonomy, while the absence of 34 issue types highlights the current limitations.

Secondly, regarding data types, the datasets encompass a diverse range, including APK, Images, Video, and Web. Notably, APK-type datasets are the most prevalent, with four in total (TSO, CCF, SMM, SAL). Image-type datasets feature HHM and MotorCheck, the latter of which also incorporates XML files associated with images. Additionally, there are three Video-type datasets (YouDescribe, AudioSet, HowTo100M) and one Web-type dataset (BM). Regrettably, none of these 10 datasets incorporate two or more data types, a limitation that may also contribute to the inability of current tools to perform cross-platform detection.

Finally, in terms of instance quantities, there is considerable variation across datasets. Video-type datasets have the highest instance totals, with 3,000,020 instances overall, led by AudioSet with 2,000,000 instances, while YouDescribe has only 20. Image-type datasets (HHM and MotorCheck) total 2,099 instances. The APK-type datasets have a combined total of 2,595 instances, with CCF leading with 2,270 instances, and SAL with only 20. The Web-type dataset BM contains only 30 instances, revealing a notable shortage of web accessibility data samples. Furthermore, all datasets except MotorCheck have been used in only one paper, indicating they are often utilized in a single study. This may be due to limited issue type coverage, which reduces broader applicability and generalizability.

In summary, datasets for detection tools collectively cover 21 of the 55 accessibility issue types, achieving a coverage rate of only 38.1\%, while 34 types remain completely uncovered, highlighting limitations in coverage. Furthermore, the datasets’ single data type and limited generalizability further indicate the limitations of existing datasets. These findings underscore the necessity for developing datasets with varied data types and broader accessibility issue type coverage.

\subsubsection{Status of Datasets for Repair Tools}
We provides a detailed analysis of 8 datasets used in repair tools, including ZCF, ACH, XLL, ZZG-1, LabelDroid-ex, among others, focusing on coverage of issue types, data types, and instance quantities.

Firstly, regarding coverage of issue types, these 8 datasets collectively cover 7 out of the 55 accessibility issue types in the taxonomy, with an overall coverage rate of 12.7\%. When considered individually, these datasets cover between 1 and 4 accessibility issue types, in contrast to the broader range (2-10 types) seen in detection tool datasets. Specifically, the ZLC dataset covers the most (4 types), while ACH, ZZG-1, LabelDroid-ex, LabelDroid, and ZZG-2 each cover only 1 type. This shows relatively low coverage for repair tool datasets in terms of our cataloged 55 issue types. Notably, issue type 24 (“Lack of Labels”) appears most frequently, covered in ZZG-1, LabelDroid-ex, LabelDroid, and ZZG-2, reflecting its prominence in accessibility repair research. Overall, these datasets cover only 7 of the 55 issue types, with 48 types entirely absent, underscoring the limitations of current dataset.

Secondly, in terms of data types, datasets for repair tools span three forms: APK, Images, and Web. Image-type datasets are the most numerous, with five (XLL, ZZG-1, LabelDroid-ex, LabelDroid, and ZLC, the ZLC includes XML files of images). APK-type datasets include two (ZCF and ACH), while the Web-type dataset includes only one (ZZG-2). Unfortunately, none of these 8 datasets contain two or more data types, limiting support for cross-platform repair tools. This limitation affects the versatility of current repair tools across multiple platforms.

Finally, regarding instance quantities, results show that Image-type datasets contain the most instances (61,654 in total), with LabelDroid-ex having the largest count at 21,864 instances, while XLL has only 31. The two APK-type datasets (ZCF and ACH) have a total of 10,126 instances, with ZCF leading at 10,078. The Web-type dataset ZZG-2 contains only 30 instances. In total, Web-type datasets are limited in both quantity and instance count, reflecting the shortage in web accessibility data sampling. Furthermore, each of these 8 datasets has been cited in only one paper, suggesting limited coverage of issue types and general applicability.

In summary, datasets for repair tools cover only 7 out of 55 accessibility issue types, achieving a coverage rate of 12.7\%, with 48 types completely missing. The single data type focus and limited general applicability further underscore the limitations of existing datasets. These findings highlight the critical need to develop datasets with diverse data types and broader accessibility issue coverage.

\begin{tcolorbox}[colback=gray!10,
			colframe=black,
			width=9cm,
			arc=2mm, auto outer arc]		
			\textbf{Answer to RQ3}: Our analysis of the 18 datasets found that the 10 datasets used in detection tools cover 21 of the 55 types in the taxonomy, achieving a coverage rate of 38.1\%, while the 8 datasets used in repair tools cover only 7 out of the 55 issue types, with a coverage rate of 12.7\%. Additionally, the datasets used in both detection and repair tools are characterized by single data types and limited general applicability. These findings underscore the crucial importance of constructing datasets with diverse data types and broader coverage of accessibility issue types.
\end{tcolorbox}

\section{Discussion}
This sector discusses the significance of the study, threats to its validity, and identifies challenges.

\subsection{Takeaway}
Our research aims to build a comprehensive taxonomy of accessibility issue types and analyze detection tools, repair tools, and related datasets based on this taxonomy, with significant implications for both academia and industry:

\begin{itemize}
\item \textbf{Standardization}:For researchers, developers, and designers, the taxonomy provides a standardized framework for classifying accessibility issues, helping reduce interpretive biases and fostering improved communication among stakeholders.
\item \textbf{Impact Assessment of Accessibility Issues}:For developers, product managers, and testers, this taxonomy clarifies known accessibility issue types and the affected user groups, aiding efforts to enhance accessibility and user experience in product design and development.
\item \textbf{Enhancement of the WCAG 2.1 Guidelines}: For researchers and standards bodies, our taxonomy highlights the accessibility issues that are not yet addressed in WCAG 2.1, encouraging further guideline expansion to cover these gaps.
\item \textbf{Optimization of Detection and Repair Tools}:For developers, researchers, and testers, our systematic evaluation of existing tools provides insights into their strengths and limitations, supporting efforts to create more comprehensive tools with broader coverage.
\item \textbf{Improvement of Datasets for Detection and Repair}:Improvement of Datasets for Detection and Repair: For developers and researchers, our dataset analysis reveals current limitations, particularly regarding issue type coverage, offering guidance for building more diverse datasets to support comprehensive detection and repair tools.
\end{itemize}
In summary, this study provides a systematic taxonomy, tool assessment, and dataset analysis that advances research and practice in accessibility. These outcomes contribute not only to academic discourse but also offer valuable insights for practitioners focused on accessibility issue detection and repair.

\subsection{Threats to Validity}
This section examines internal and external threats to the validity of this research.
\subsubsection{Internal Threats}
The study’s internal validity may be affected by limitations in literature selection and potential subjectivity in constructing the taxonomy, as summarized as follows:
\begin{itemize}
\item \textbf{Limitations in Literature Selection}:Our literature review concluded in June 2024, excluding any subsequent studies. This date restriction may have led to the omission of recent advancements, potentially introducing selection bias and affecting comprehensiveness.
\item \textbf{Subjectivity in Taxonomy Construction}:The manual merging and classification of accessibility issues based on existing literature and guidelines could introduce subjective judgments, possibly affecting the accuracy of both the merging and classification outcomes.
\end{itemize}

\subsubsection{External Threats}

External validity threats include the emergence of new accessibility issues and limitations in the survey process, as summarized as follows:
\begin{itemize}
\item \textbf{Emergence of New Issues}:With the rapid development of technology, new accessibility issues may arise from the introduction of novel interaction devices and platforms, potentially limiting the taxonomy's applicability. To address this, we have developed and maintained a dedicated website for ongoing updates.
\item \textbf{Survey Limitations}:Our validation survey may not fully capture perspectives across diverse cultural backgrounds, as individuals from different cultures may have varying understandings of and expectations for accessibility. Future work will aim to broaden participation, especially from accessibility professionals worldwide, to gather more comprehensive feedback.
\end{itemize}

\subsection{Challenges}
Building on this study, we identify seven challenges in accessibility issue research.
\begin{itemize}
\item \textbf{Dynamic Taxonomy Updates}: With advancements in technologies such as VR, AR, smart devices, and IoT, new accessibility issues are emerging rapidly, presenting unique challenges that traditional taxonomies may not fully capture. Future research may need to develop a dynamic taxonomy that can evolve with new technologies.

\item \textbf{Limitations of Detection Tools}: Current detection tools are limited in coverage, with many identifying only 1–4 issue types, leaving 24 types undetected. This limits developers’ ability to fully address accessibility issues in diverse scenarios. Future research should aim to create more comprehensive detection tools that can operate across a wider range of contexts.
\item \textbf{Limitations of Detection and Repair Tools}: Current detection tools are limited, identifying only 1-4 issue types, leaving 24 undetected. Repair tools address even fewer types, leaving 42 unresolved. This limits developers' ability to address accessibility issue types comprehensively. Future research should develop more comprehensive tools covering a wider range of issue types.
\item \textbf{Cross-Platform Compatibility}: Many accessibility tools and datasets focus on specific platforms (e.g., mobile or web) without supporting cross-platform functionality, resulting in inconsistent accessibility experiences across devices. Future work should explore developing tools that function seamlessly across multiple platforms.
\item \textbf{Lack of a Standardized Evaluation Framework}: Existing tools use diverse evaluation metrics, such as precision, recall, and success rates, making direct comparisons challenging. A standardized framework is needed to unify performance evaluation, aiding researchers and practitioners in tool selection.
\item \textbf{Limited Public Access to Tools and Datasets}: Our study revealed that some tools and datasets are not publicly accessible, restricting comprehensive evaluation. Increased public access would enable researchers to verify and improve these resources.
\end{itemize}

In summary, these challenges outline clear directions for future research in accessibility, emphasizing the need for dynamic taxonomies, robust tools, comprehensive datasets, and standardized evaluation frameworks to enhance detection and repair capabilities and promote inclusive digital environments.
\section{Conclusion}
This study addresses gaps in the accessibility research landscape by constructing a taxonomy of accessibility issue types, evaluating detection and repair tools, and assessing the current state of available datasets. Through a comprehensive analysis of 42 studies, we developed a taxonomy encompassing 55 different types of accessibility issues. This taxonomy categorizes each issue type and annotates WCAG 2.1 guidelines, affected user groups, and application scenarios, helping developers, designers, and stakeholders better understand accessibility issues and fostering more inclusive design practices. Social survey feedback also confirmed broad support for this taxonomy.

Additionally, based on the taxonomy, we conducted a comprehensive evaluation of existing tools and datasets, uncovering limitations in issue detection, repair capabilities, and dataset coverage. Overall, this study provides a more systematic understanding of real-world accessibility issues, offering valuable insights for future research and practice. We hope these findings inspire further exploration, drive advancements in accessibility technologies, and contribute to building a more inclusive and accessible digital society.

\bibliography{references}

\end{document}